\documentclass[12pt]{article}
\begin{document}

\voffset -0.6 true cm
\hoffset 1.3 true cm
\topmargin 0.0in
\evensidemargin 0.0in
\oddsidemargin 0.0in
\textheight 8.6in
\textwidth 6.9in
\parskip 10 pt

\newcommand{\be}{\begin{equation}}
\newcommand{\ee}{\end{equation}}
\newcommand{\bea}{\begin{eqnarray}}
\newcommand{\eea}{\end{eqnarray}}
\newcommand{\beas}{\begin{eqnarray*}}
\newcommand{\eeas}{\end{eqnarray*}}

\newcommand{\ms}{m_{\rm susy}}
\newcommand{\mpl}{m_{\rm planck}}

\let\a=\alpha\let\b=\beta\let\d=\delta
\let\e=\epsilon\let\f=\phi\let\g=\gamma
\let\h=\eta\let\th=\theta\let\k=\kappa\let\l=\lambda
\let\m=\mu\let\n=\nu\let\p=\pi\let\r=\rho
\let\s=\sigma\let\t=\tau\let\u=\upsilon
\let\w=\omega\let\x=\xi\let\y=\psi
\let\z=\zeta\let\G=\Gamma\let\P=\Pi\let\S=\Sigma
\let\Th=\Theta\let\D=\Delta\let\F=\Phi
\let \L =\Lambda

\begin{titlepage}
\begin{flushright}
{\small CU-TP-1006} \\
{\small RUNHETC-2001-06} \\
{\small hep-ph/0102309}
\end{flushright}

\begin{center}

\vspace{1cm}

{\Large \bf Testing cosmological supersymmetry breaking}

\vspace{1cm}

Daniel Kabat ${}^1$ and Arvind Rajaraman ${}^2$

\vspace{4mm}

${}^1${\small \em Department of Physics, Columbia University} \\
{\small \em New York, NY 10027} \\
{\small \tt kabat@physics.columbia.edu}

\vspace{2mm}

${}^2${\small \em Department of Physics, Rutgers University} \\
{\small \em Piscataway, NJ 08855--0849} \\
{\small \tt arvindra@physics.rutgers.edu}

\end{center}

\vskip 1.3 cm

\noindent
Banks has proposed a relation between the scale of supersymmetry
breaking and the cosmological constant in de Sitter space.  His
proposal has a natural extension to a general FRW cosmology, in which
the supersymmetry breaking scale is related to the Hubble parameter.
We study one consequence of such a relation, namely that coupling
constants change as the universe evolves.  We find that the most
straightforward extension of Banks' proposal is disfavored by
experimental bounds on variation of the fine structure constant.

\end{titlepage}

\section{Introduction}

The Holographic Principle \cite{holography} states that the total
number of degrees of freedom in a theory of quantum gravity scales
like the surface area.  This is radically different from the behavior
of quantum field theory, in which the number of states at high energy
scales like the volume.  This suggests that the usual field theory
calculation of divergent radiative corrections to scalar masses is
modified by quantum gravity.

Moreover, holographic theories have a UV/IR connection \cite{UVIR},
which relates high energies to long distances.  In particular the
spectrum of high energy states in a holographic theory may well be
determined by the large scale structure of the universe.  Putting
these ideas together, it seems plausible that in a holographic theory,
scalar masses are related to cosmology.

Banks has put forward a very concrete proposal for such a relation
\cite{banks}.  He considers M-theory in a de Sitter background, with a
cosmological constant $\Lambda > 0$ corresponding to a vacuum energy
density
\[
\rho_{\rm vac} = m_{\rm vac}^4 = {\Lambda \over 8 \pi G}\,.
\]
The de Sitter geometry breaks supersymmetry.  Banks proposes that the
supersymmetry breaking scale is not given by the naive guess $\ms
\approx m_{\rm vac}$.  Rather, he suggests that UV/IR effects could
enhance this to
\[
\ms \approx \left(\mpl m_{\rm vac}\right)^{1/2}\,.
\]
With a Planck mass of $10^{19}\,GeV$ and a vacuum energy of
$10^{-3}\,eV$, this leads to a phenomenologically acceptable breaking
of SUSY at the few $TeV$ scale.

In this paper we generalize Banks' proposal to a general flat FRW
cosmology (section 2).  The natural generalization relates the SUSY
breaking scale to the Hubble parameter.  Via the renormalization group
a change in the SUSY breaking scale affects low energy coupling
constants, so coupling constants will change as the universe evolves.
We consider the experimental bounds on variation of the fine structure
constant in section 3, and show that the simplest extension of Banks'
proposal is experimentally disfavored.  Section 4 contains our
conclusions.

\section{Cosmological supersymmetry breaking in an FRW universe}

Banks has proposed that in a de Sitter background the scale of
supersymmetry breaking is set by
\[
\ms \approx \mpl^{1/2} \left(\Lambda \over 8 \pi G \right)^{1/8}\,.
\]
This formula is supposed to be a consequence of the finite number of
states which are available in a de Sitter background.  This motivates
us to begin our search for an appropriate generalization of Banks'
formula by rewriting $\ms$ in terms of the entropy of de Sitter space
$S = 3 \pi /G \Lambda$.
\be
\ms \approx \mpl^{1/2}  \left( {3 \over 8 G^2 S} \right)^{1/8}
\label{susyrelation}
\ee
This relation has a natural extension to a general spacetime: {\em the
supersymmetry breaking scale is given by (\ref{susyrelation}), with
$S$ interpreted as the holographic bound on the cosmological entropy}.

For the remainder of this paper, we specialize to flat ($k = 0$)
FRW universes, with metric
\[
ds^2 = - dt^2 + R^2(t) \left(dr^2 + r^2 d\Omega_2^2\right)
\]
and Hubble parameter $H = \dot{R}/R$.  Following \cite{Bousso, CosmoHolo}
we take the cosmological entropy to be bounded by the area of the apparent
horizon.  The apparent horizon is a sphere of radius $r_{AH} = 1/HR$ and
area $A_{AH} = 4 \pi / H^2$, corresponding to an entropy
\be
S = {A_{AH} \over 4 G} = {\pi \over G H^2}\,.
\label{entropy}
\ee
Thus the SUSY breaking scale is related to the Hubble parameter by
\be
\ms \approx \mpl^{1/2} \left( {3 H^2 \over 8 \pi G} \right)^{1/8}\,.
\label{susyrelation1}
\ee
Curiously, the quantity in parenthesis is the critical density.  For
our purposes, it is more convenient to rewrite this as a ratio, with a
subscript $0$ denoting the present.
\be
{\ms(t) \over \ms(t_0)} = \left( {H(t) \over H_0} \right)^{1/4}
\label{susyrelation2}
\ee
This implies that the supersymmetry breaking scale changes with time
as the universe evolves.

Although the scale of supersymmetry breaking has not been directly
observed, we can put limits on any possible changes in $\ms$ because
the scale of supersymmetry breaking affects the values of the
low-energy coupling constants.  This dependence on $\ms$ arises from
the renormalization group, which states that at one loop gauge
couplings $\alpha \equiv g^2 / 4 \pi$ evolve with scale according to
\[
{1 \over \alpha(\mu^2)} - {1 \over \alpha(M^2)} = {b_0 \over 2 \pi}
\log {\mu \over M}\,.
\]
The $\beta$-functions generally change at the scale where
supersymmetry is broken, and this makes the low-energy couplings
sensitive to the value of $\ms$.

The precise dependence on $\ms$ is easily obtained.  The Hubble
parameter has been decreasing as the universe evolves, so the
supersymmetry breaking scale (\ref{susyrelation1}) has been decreasing
with time.  Suppose that in the course of this evolution the
supersymmetry breaking scale drops from an initial value $m_1$ at time
$t_1$, to a new value $m_2$ at time $t_2$.  Above the scale $m_1$ we
assume that couplings are not affected by cosmology, so the
couplings are identical at the two different times:
\[
\alpha(m^2)\vert_{t_1} = \alpha(m^2)\vert_{t_2} \qquad \hbox{\rm for $m > m_1\,.$}
\]
We also assume that below the scale $m_2$ the runnings are the same:
\[
b_0\vert_{t_1} = b_0\vert_{t_2} \qquad \hbox{\rm below the scale $m_2$}\,.
\]
However in the intermediate range $m_2 < m < m_1$ the
$\beta$-functions are different at the two different times.  This
leads to a change in the value of the observed couplings at low
energy, which is given by the difference in the two $\beta$-functions.
\be
\left.{1 \over \alpha}\right\vert_{t_2} - \left.{1 \over \alpha}\right\vert_{t_1} =
{1 \over 2 \pi} \left(b_0^{SM} - b_0^{MSSM}\right) \log {m_1 \over m_2}
\label{diff}
\ee

We now specialize to the evolution of the fine structure constant.  In
the standard model the photon is a mixture of the $U(1)_Y$ and
$SU(2)_L$ gauge fields, with coupling
\[
{1 \over \a} = {1\over \a_Y} + {1 \over \a_{SU(2)}}\,.
\]
The appropriate one-loop beta functions in the standard model and its
supersymmetric extension are therefore given by \cite{Weinberg}
\be
b_0 = b_0^{Y} + b_0^{SU(2)} =
\left\lbrace
\begin{array}{cl}
-10/3 & \qquad \hbox{\rm standard model} \\
-12   & \qquad \hbox{\rm MSSM}
\end{array}
\right.
\ee
This implies that the fine structure constant depends on time
according to
\[
{1 \over \alpha(t_0)} - {1 \over \alpha(t)} =
{13 \over 3 \pi} \log {\ms(t) \over \ms(t_0)}\,.
\]
Given the proposed relationship (\ref{susyrelation2}) between $\ms$
and the Hubble parameter, this implies that
\be
{1 \over \alpha(t_0)} - {1 \over \alpha(t)} =
{13 \over 12 \pi} \log {H \over H_0}\,.
\label{result}
\ee
Since $H$ was larger in the past, our proposal implies that $\alpha$
was larger in the past.

The evolution of the Hubble parameter is determined by the Friedmann
equation
\[
H^2 = {8 \pi G \over 3} \sum_i \rho_i
\]
where $\rho_i$ are the various components of the energy density.  We
will model the universe as dominated by matter plus vacuum energy.
This leads to an equation for the evolution of the normalized scale
factor $a(t) = R(t)/R_0$,
\be
\left({\dot{a} \over a}\right)^2 = H_0^2 \left(\Omega_\Lambda + \Omega_M a^{-3}\right)
\label{adotOvera}
\ee
where $\Omega_\Lambda$, $\Omega_M$ are the present-day fractions of
the critical density.  Thus the Hubble parameter is given by
\be
{H \over H_0} = \sqrt{\Omega_\Lambda + {\Omega_M \over a^3}}\,.
\label{Hubble}
\ee
We also have a relation between the scale factor and the age of the
universe,
\be
t = {1 \over H_0} f(a)
\label{age}
\ee
where
\[
f(a) = {2 \over 3 \sqrt{\Omega_\Lambda}} \log \left[
\,\left({\Omega_\Lambda a^3 \over \Omega_M}\right)^{1/2} +
\sqrt{1 + {\Omega_\Lambda a^3 \over \Omega_M}}\,\,\right]\,.
\]

\section{Experimental bounds}

Experimental constraints on the time variation of $\alpha$ come from a
variety of sources.  Direct lab measurements were performed by
\cite{Prestage} using clocks based on ultra-stable atomic oscillators.
By comparing rates from different clocks, the authors obtained a bound
\be
\vert \dot{\a}/\a \vert \le 3.7 \times 10^{-14} \ yr{}^{-1}\,.
\label{LabBound}
\ee
Equations (\ref{result}) and (\ref{Hubble}) predict that at the present time
\bea
\dot{\a}/\a & = & - {13 \over 8 \pi} \, \alpha \, \Omega_M H_0 \nonumber \\
& = & (- 8.7 \pm 1.8) \times 10^{-14} \ yr{}^{-1}
\label{present}
\eea
where we have used the values $\Omega_M = 0.3$, $H_0 = h / (9.78 \times
10^9 \, yr)$ and the uncertainty corresponds to varying $h$ from 0.6 to
0.9 \cite{Overduin}.  This is about a factor of two larger than the
experimental bound (\ref{LabBound}).

A much more stringent bound comes from the Oklo reactor, a natural
nuclear reactor which was triggered about 1.8 billion years ago.  From
an analysis of this phenomenon, the authors of \cite{Damour:1996zw}
obtained a bound
\be
-0.9 \times 10^{-7} < {\Delta \alpha \over \alpha} < 1.2 \times 10^{-7}\,.
\ee
Equations (\ref{result}), (\ref{Hubble}), (\ref{age}) predict an
effect which is three orders of magnitude larger:
\[
{\Delta \alpha \over \alpha} \equiv {\alpha(t) - \alpha_0 \over \alpha_0}
= (1.9 \pm 0.4) \times 10^{-4}\,.
\]
Again the quoted uncertainty corresponds to $\Omega_M = 0.3$ and
$h$ ranging from 0.6 to 0.9.  This seems like a disaster for our
proposal (\ref{susyrelation2}), but as we discuss in the conclusions,
this result should be qualified.

Another bound comes from observation of quasar absorption lines.  One
group has obtained the bound \cite{quasar1}
\be
{\Delta \alpha \over \alpha} = (-4.6 \pm 4.3 \pm 1.4) \times 10^{-5}
\label{quasarbound}
\ee
at redshifts $z \sim$ 2 -- 3, while other groups have reported
stronger results \cite{quasar2}.  Equations (\ref{result}) and
(\ref{Hubble}) give
\[
{\Delta \alpha \over \alpha} = 2.1 \times 10^{-3}
\]
at $z = 1.5$ (recall $a = 1/(1+z)$).  Even compared to the
conservative bound (\ref{quasarbound}), this is about two orders of
magnitude too large.

Finally, we consider the bound from big bang nucleosynthesis.  A
recent bound was obtained by \cite{BBN}, who found a limit
\[
{\Delta \alpha \over \alpha} = (-7 \pm 9) \times 10^{-3}\,.
\]
At the time of nucleosynthesis the universe was radiation-dominated,
with $H(t) = 1/2t$.  Equation (\ref{result}) predicts that at the
end of nucleosynthesis
\be
{\Delta \alpha \over \alpha} = (9.7 \pm 0.1) \times 10^{-2}
\label{BBNresult}
\ee
where we have set $t_{BBN} = 100 \, {\rm sec}$.  This is about an
order of magnitude larger than the experimental bound.

One might wonder whether we are allowed to apply our formulas to the
early universe, since the relation (\ref{susyrelation}) is only
expected to be valid for very large values of the de Sitter entropy
\cite{banks}.  In a radiation dominated universe the entropy follows
from (\ref{entropy}), $S = 4 \pi t^2 / t_{\rm planck}^2$.  Thus $S
\approx 10^{90}$ at the end of nucleosynthesis; presumably this is
large enough for our generalization of (\ref{susyrelation}) to hold.

\section{Conclusions}

In this paper we have presented an extension of Banks' cosmological
supersymmetry breaking proposal to a flat FRW universe.  As we have
seen, bounds on variation of the fine structure constant provide a
stringent test of this extended proposal: it seems to be ruled out,
especially by the Oklo reactor data.

We reached this conclusion by studying the behavior of the fine
structure constant, while neglecting the cosmological evolution of all
other parameters in the standard model.  In a sense this makes our
analysis very conservative, since one generally expects that relevant
couplings in the standard model should have a power-law dependence on
$m_{SUSY}$.  Given the proposal (\ref{susyrelation2}), this would give the
relevant couplings a power-law dependence on the Hubble parameter, in
gross contradiction with experiment.  However, the behavior of
relevant couplings may depend on the details of the way in which
supersymmetry breaking is communicated to the standard model.
  
We seem to have found that the proposal (\ref{susyrelation2}) can be ruled
out just by considering marginal operators.  However, this conclusion
should be qualified, because there are some potentially important
effects at the level of marginal operators that we have ignored.  For
example, we have neglected the fact that according to our proposal
cosmological evolution should also affect the QCD scale.  On the one
hand, changes in the QCD scale should be tightly constrained by big
bang nucleosynthesis.  But on the other hand changing the QCD scale
may well modify the analysis \cite{Damour:1996zw} of the Oklo reactor
data, which implicitly assumed that the QCD scale was constant.

To address this concern, let us note that ref.~\cite{Sisterna:1990et}
carried out a global fit to numerous observations (including a much
more conservative analysis of the Oklo data than
\cite{Damour:1996zw}).  They allowed $\Lambda_{QCD}$, $G_F$, $\alpha$,
$G_N$ and $m_e$ to vary independently, and found an upper bound
\[
\vert \dot{\alpha} / \alpha \vert < 1.4 \times 10^{-15} \ yr{}^{-1}
\qquad \hbox{(95\% confidence level)\,.}
\]
This limit is almost two orders of magnitude smaller that our
present-day prediction (\ref{present}).

It is possible that the supersymmetry breaking scale is
determined by cosmology, but not in the way that we have suggested.
We took the entropy that appears in (\ref{susyrelation}) to be given
by the area of the apparent horizon.  This seems quite natural,
following \cite{Bousso}, but other choices could be contemplated, such
as the area of the event horizon.  It could also be that the entropy
is determined by the value of $\Lambda$, even if $\Lambda$ never plays
an important role in the evolution of the universe.
Ref.~\cite{Bousso2} provides some support for this possibility.

Of course, it could be that supersymmetry breaking and cosmology are
unrelated, or that such a relation exists but is far more subtle than
anything we have discussed here.

\bigskip
\centerline{\bf Acknowledgments}
\noindent
We are grateful to Bobby Acharya, Tom Banks, Michael Graesser, Lam
Hui, Norihiro Iizuka and Scott Thomas for valuable discussions.  DK
wishes to thank Rutgers University for hospitality during the course
of this work.  DK is supported by the DOE under contract
DE-FG02-92ER40699.  AR is supported in part by DOE grant
DE-FG02-96ER40559.


\end{document}